\documentclass[twocolumn]{elsart}

\usepackage{natbib}

\usepackage{graphicx}

\usepackage{amssymb}

\begin{document}

\begin{frontmatter}



\title{X-ray Emission Processes in Extragalactic Jets, Lobes and Hot Spots}


\author{Andrew S. Wilson}

\address{Astronomy Department, University of Maryland, College Park, MD 20742}
\ead{wilson@astro.umd.edu}
\begin{abstract}

This paper is a brief review of the processes responsible for X-ray emission
from radio jets, lobes and hot spots. Possible photons in inverse
Compton scattering models include the radio synchrotron radiation itself
(i.e. synchrotron self-Compton [SSC] emission), the cosmic microwave background 
(CMB),
the galaxy starlight and radiation from the active nucleus. 
SSC emission has been detected from a number of hot spots.
Scattering of the 
CMB is expected to dominate for jets (and possibly hot spots) undergoing 
bulk relativistic motion close to the direction towards the
observer. Scattering of
infrared radiation from the AGN should be observable from radio lobes, 
especially if they are close to the active nucleus.
Synchrotron radiation is detected in some sources, most
notably the jet of M87. I briefly discuss why different hot
spots emit X-rays by different emission mechanisms and the nature of the 
synchrotron spectra.

\end{abstract}

\begin{keyword}
galaxies:active \sep galaxies:jets \sep galaxies:nuclei \sep magnetic fields
\sep X-rays:galaxies


\end{keyword}

\end{frontmatter}

 
\section{Introduction}

X-ray emission from diffuse cosmic gases may result from thermal or 
non-thermal processes. An X-ray emitting thermal gas can be collisionally 
ionized, in which case the temperature needs to be $\simeq$ 10$^{7-8}$K, or 
photoionized, when the temperature is much lower (several $\times$ 10$^{4}$K
- several $\times$ 10$^{5}$K). Chandra and XMM observations have shown that the
latter type of gas is responsible for the X-ray emission of several bright
Seyfert galaxies. For example, in NGC 1068 the widths of the radiative 
recombination continua reveal that the temperature is only a few
$\times$ 10$^{4}$K, indicating the gas is photoionized (and photoexcited) with 
a range of 
ionization parameters (Kinkhabwala et al. 2002; Brinkman et al. 2002). 
The non-thermal processes are inverse Compton (IC) scattering
and synchrotron radiation. The energies of the relativistic electrons tend to
be low if the radiation results from IC emission (Section 2) and
high if it results from synchrotron radiation ($\gamma$ = E/mc$^{2}$ $\simeq$
10$^{7}$ - 10$^{8}$ for radiation in the Chandra band and a canonical 
B = 10$^{-4}$ gauss magnetic field strength).

Thermal X-ray emission has not, to my knowledge, been detected from any
extragalactic jet. Detection of IC
emission allows the magnetic 
field strength to be measured directly as long as the relevant photon density 
is known and there is no bulk relativistic motion. If the X-rays are 
synchrotron, we infer the presence of high energy electrons and/or positrons
with synchrotron half lives or order years, in which case X-ray images
highlight regions of current particle acceleration.

Bulk relativistic motion
is described by the Doppler parameter
$\delta = [L(1\ - \beta\ \cos\theta)]^{-1}$, where
$L$ is the Lorentz factor of the bulk flow, $\beta$ is the
bulk velocity in units of the speed of light, and $\theta$ is
the angle between the velocity vector and the line of sight. If
S$_{\nu}^{'}$ $\propto$ $\nu^{-\alpha}$ is the flux density which would
be observed in the absence of bulk motion and the ``blob''
emits isotropically in its own
frame, the observed spectrum
S$_{\nu}$ $\propto$ $\delta^{3 + \alpha}$S$_{\nu}^{'}$ (e.g. Begelman et al.
1984). This dependence
on $\delta$ is 
usually assumed for synchrotron and synchrotron self-Compton (SSC) emission. 
On the other hand, if the radiation being scattered is isotropic in the
Hubble (or galaxy) frame (so called external Compton [EC] emission), then 
the principal
dependence of the observed spectrum is
S$_{\nu}$ $\propto$ $\delta^{4 + 2\alpha}$S$_{\nu}^{'}$ (Dermer 1995).
The magnetic field derived from SSC or EC models is then
a function of $\delta$, the function being different in the two cases
(e.g. Tavecchio et al. 2000; Celotti et al. 2001). 

\section{Inverse Compton Scattering - Sources of Photons}

A  key feature of IC models is that the spectral index of the
X-ray emission should be the same as that of the synchrotron emission from
the same population of electrons. Thus measurement of the X-ray spectral index
is vital for confirmation of IC models. However, the synchrotron
emission may be at too low a frequency to be observable.
Possible sources of photons in IC models include:

\noindent
a) the synchrotron radiation itself - i.e. SSC emission.

\noindent
b) the cosmic microwave background radiation (CMB). To scatter this radiation
into the Chandra band requires $\gamma$ $\simeq$ 10$^{3}$. Such electrons
radiate synchrotron radiation at meter wavelengths. As pointed out by Schwartz
(2002), the enhancement of the CMB energy density by a factor of 
(1 + z)$^{4}$ compensates for the (1 + z)$^{-4}$ decrease of surface
brightness with increasing redshift, z, and X-ray emission from such 
relativistic jets pointed towards us should be seen to 
large distances in the universe. 

\noindent
c) the galaxy starlight. Scattering into the Chandra band would require very
low energy ($\gamma$ $\simeq$ 20) cosmic rays. Such electrons would radiate
synchrotron radiation at $\simeq$ 0.1 MHz for our canonical 
B = 10$^{-4}$ gauss, too low a frequency to be observable.

\noindent
d) radiation from the active nucleus, which may be ``beamed'' along the radio
axis. The radiation density is probably dominated by infrared emission, 
so we require $\gamma$ $\simeq$ 100 - 300, which would radiate synchrotron 
radiation at a few to tens of MHz. 

In the following, I provide examples of each of the different radiation
processes.

\section{Inverse Compton Scattering}
\subsection{Synchrotron Self-Compton Emission}

Chandra has detected all four of the radio hot spots (A, B, D and E) in Cygnus
A and obtained spectra for the two brightest ones (A and D). Both are well
described by a power law with spectral index $\alpha$ = 0.8 $\pm$ 0.2 absorbed
by the Galactic column in the direction of Cygnus A (Wilson et al. 2000).
Thermal X-ray models require too high gas densities and may be ruled out. The
images and spectra strongly support SSC models of the
X-ray emission, as proposed by Harris et al. (1994) on the basis of ROSAT 
imaging observations. Such models indicate that the magnetic field in each of 
the brighter hot spots is 1.5 $\times$ 10$^{-4}$ gauss, with an uncertainty of
a few tens of percent. This field strength is close to 
equipartition. A few other cases of SSC emission from hot spots are
known (e.g. 3C 295 - Harris et al. 2000; 3C 123 - Hardcastle et al. 2001) and
the number is likely to increase dramatically with further Chandra
observations.

In order to check the predictions of the SSC model,
P. L. Shopbell and I have observed the hot spots of Cygnus A with HST and
found evidence for diffuse optical emission on the northern edge of hot spot D.
The apparent displacement of the optical emission from the peak of the radio
and X-ray emissions may result from astrometric uncertainties in the HST data.
The strength of the optical emission is consistent with, though somewhat higher
than, the predictions of the SSC model (Fig. 1). However, this optical
emission needs to be confirmed by deeper HST observations.

\begin{figure}
\includegraphics[scale=0.35]{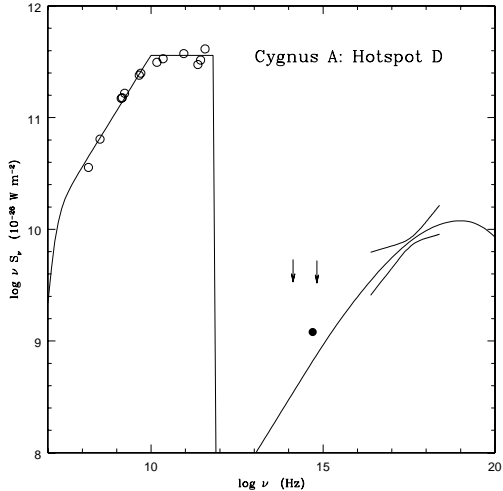}
\caption{The spectrum of hot spot D in Cygnus A (adapted from Wilson et al.
2000). The pentagons represent radio measurements, the two downward pointing 
arrows
represents upper limits to the optical and near infrared emissions from 
ground-based measurements (Meisenheimer et al. 1997), the filled circle
represents a tentative detection (requiring confirmation) with HST at 
$\simeq$ 6000 \AA\ and the bow tie is the Chandra-measured boundary of the
X-ray spectrum (90\% confidence). The continuous line is a representation of the
radio spectrum and the predicted SSC spectrum for 
B = 1.5 $\times$ 10$^{-4}$ gauss.}
\label{fig:fig1}
\end{figure}

\subsection{IC Scattering of the CMB}

The quasar PKS0637-752 was the first celestial target of Chandra. The 
observations revealed an X-ray jet coincident with the inner portion of the
jet previously detected in the radio (Schwartz et al. 2000). Schwartz et al.
(2000) noted that an SSC model implies extreme departures from the conventional
minimum energy criterion. Tavecchio et al. (2000) and Celotti et al. (2001)
showed that IC scattering of the CMB is consistent with equipartition for a
Doppler factor $\delta$ $\simeq$ 10. This process is plausible since VLBI
observations reveal superluminal motion on pc scales, implying $\theta$
$<$ 6$^{\circ}$. This interpretation implies that the bulk motion of the jet is
relativistic at nuclear distances up to several hundred kpc, a result consistent
with systematic polarisation asymmetries in powerful radio galaxies
and quasars (Laing 1988; Garrington et al.
1988).

We independently evaluated this process for the X-ray emission of the 100 kpc 
long jet of Pictor A (Wilson et al. 2001). There it was found that an
equipartition model requires $\theta$ $\simeq$ 8$^{\circ}$, which is too small
for this lobe dominated radio galaxy. For more plausible values of $\theta$,
the magnetic field would be below equipartition.

\subsection{IC Scattering of Nuclear Photons}

In the unified model of AGN, the nucleus is surrounded by a disk or torus of
gas and dust. This structure collimates much of the radiation into two wide
cones along and around the radio axis. Thus the lobes of radio galaxies are 
expected to be illuminated by intense radiation from a ``hidden'' quasar.
Additionally, there may be a much narrower ``blazar'' beam emitted by a
relativistic jet in the nucleus.
Relativistic electrons with $\gamma$ $\sim$ 100 - 300 will IC scatter infrared
photons from the quasar into the Chandra band. This process should dominate
scattering of the CMB for sources whose axis is not close to the line of sight,
for radio lobes/jets with only mild or non-relativistic bulk motion and for
radio lobes/jets close to the nucleus. 

Brunetti et al. (1997) have calculated the anisotropic IC scattered emission
for this situation. 
Because of the angle dependence of the
IC scattering, the more distant lobe is predicted to be somewhat
brighter, depending on the value of $\theta$.
Brunetti et al. (2001, 2002)
have argued that diffuse X-ray emission from radio lobes in 3C 295 
and 3C 207 is produced by this process.

\section{Synchrotron radiation}

The strongest case for synchrotron X-ray emission in an extragalactic jet is
M87. In this jet, the X-ray spectra of all the individual knots are steeper
than their radio or optical spectra (Marshall et al. 2002; Wilson \& Yang 2002),
which argues against IC models. Further, knot HST-1 is highly variable in 
X-rays, strongly favoring synchrotron radiation (Harris, these proceedings;
Harris et al. 2003). A detailed analysis of the X-ray emission of the M87 jet
is given by Perlman \& Wilson (these proceedings and in preparation). Harris
\& Krawczynski (2002) present a list of other possible X-ray synchrotron
sources.

\section{Problems}
\subsection{Why do some hot spots show SSC and others synchrotron X-rays?}

A promising suggestion has been made by Georganopoulos \& Kazanas (this
conference). They propose that all hot spots are essentially the same, with
the X-ray and optical synchrotron emission being significantly relativistically
beamed along the source axis. The flow within the hot spots is relativistic
and decelerating. The highest energy electrons (X-ray and optical emitting)
are located in the immediate post (jet) shock region, and their beaming pattern
is narrow. Further downstream, the flow is slower with a wider beaming pattern
and these electrons emit the radio synchrotron and X-ray SSC emission. It will
be important to confirm that the SSC emitting hot spots tend to be in objects 
with axis close to the plane of the sky and that the one sided, synchrotron
emitting hot spots are viewed more pole-on.

\subsection{Synchrotron spectra}

X-ray synchrotron emitting electrons have energy half lives of order years and
must be continuously reaccelerated. In some cases, the X-ray spectrum is not a
simple extension of the radio-optical spectrum with a monotonically increasing
spectral index with increasing frequency, as would be expected for
synchrotron or IC losses. Examples of this unexpected behavior include the W
hot spot in Pic A and knots A, B (Fig. 2) and possibly D in the M87 jet. 

\begin{figure}
\includegraphics[scale=0.3, angle=90]{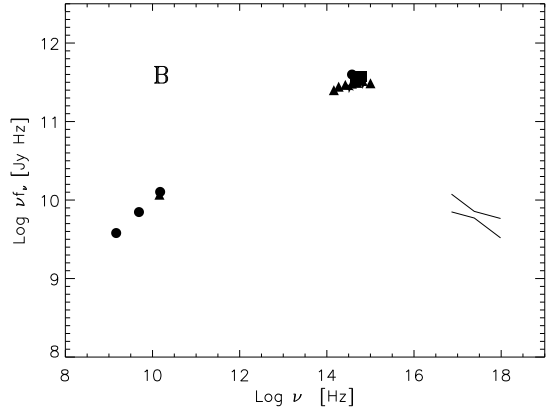}

\caption{Spectrum of knot B in the M87 jet (from Wilson \& Yang 2002)}

\label{fig:fig2}
\end{figure}

Wilson \& Yang (2002) discuss various explanations for this behavior, 
including the possibility that the X-rays come from a separate population
of electrons from those responsible for the radio and optical emissions. One
particular example of this type of model postulates that the volume within 
which the particles
are accelerated is a function of the particle energy (cf. Perlman et al. 2001;
Perlman \& Wilson in preparation). The spectrum observed at Earth would also be 
affected
if the degree of beaming is a function of particle energy 
(cf. Georganopoulos \& Kazanas, these proceedings) and hence frequency.
Lastly, Dermer \& Atoyan (2002) have shown that, when IC cooling on the CMB in
the Thomson regime exceeds synchrotron cooling, a hardening in the electron 
spectrum is formed at electron energies where Klein-Nishina effects become 
important. This effect results from the reduction in Compton losses in the
Klein-Nishina regime compared to the dE/dt $\propto$ E$^{2}$ losses of the
Thomson regime. All of these processes may serve to modify the observed shape of
the synchrotron spectrum.

This work was supported by NASA grants NAG 8-1027 and NAG 8-1755.




\end{document}